# Accelerating CALYPSO Structure Prediction by Data-driven Learning of Potential Energy Surface

Qunchao Tong,[1,2] Lantian Xue,[1] Jian Lv,[1,2,*] Yanchao Wang[1] and Yanming Ma[1,*]

[1]*State Key Laboratory of Superhard Materials, College of Physics, Jilin University, Changchun 130012, China. E-mail: mym@calypso.cn*

[2]*College of Materials Science and Engineering, Jilin University, Changchun 130012, China. E-mail: lvjian@calypso.cn*

*Ab initio* structure prediction methods have been nowadays widely used as powerful tools for structure searches and material discovery. However, they are generally restricted to small systems owing to the heavy computational cost of underlying density functional theory (DFT) calculations. In this work, by combining state-of-art machine learning (ML) potential with our in-house developed CALYPSO structure prediction method, we developed two acceleration schemes for structure prediction toward large systems, in which ML potential is pre-constructed to fully replace DFT calculations or trained in an on-the-fly manner from scratch during the structure searches. The developed schemes have been applied to medium- and large-sized boron clusters, which are challenging cases for both construction of ML potentials and extensive structure searches. Experimental structures of $B_{36}$ and $B_{40}$ clusters can be readily reproduced, and the putative global minimum structure for $B_{84}$ cluster is proposed, where substantially less computational cost by several orders of magnitude is evident if compared with full DFT-based structure searches. Our results demonstrate a viable route for structure prediction toward large systems via the combination of state-of-art structure prediction methods and ML techniques.

# 1. INTRODUCTION

The theoretical structure prediction methods are nowadays in common use and playing an increasingly important role in computational material discovery, as they can provide putative ground-state structure of materials as well as design of structures with desired functionality in aiding experimental synthesis.[1] These methods generally involve the exploration of the potential energy surface (PES) of materials through various structural sampling techniques and optimization algorithms.[2–12] We have developed a swarm-intelligence based CALYPSO structure prediction method,[13,14] which has been successfully applied to a variety of material systems.[15,16] Despite their widespread successes, structure prediction methods are inherently limited by the underlying computational workhorse of energy evaluation based on quantum mechanical methods, e.g., density functional theory (DFT), which prohibitively expensive for large systems.

Machine Learning (ML) as data-driven methods for making prediction, decision or classification have now become pervasive and deeply entrenched components of modern science and technology. They are now starting to enter the heart of computational physics, chemistry and material sciences in a manifold way.[17] Particularly, with plenty of quantum mechanical data available, various ML schemes have provided enormous accurate predictions of a rich variety of materials' properties including atomization energy,[18–22] formation energy[23–25] and various electronic properties.[20,26–28] Moreover, they have also been used to reconstruct the potential energy surface of materials based on data obtained from first-principles calculations.[29] A series of ML potentials have been developed by utilizing techniques such as artificial neural network,[30–33] Gaussian process regression,[34–36] support vector machine[37] and kernel ridge regression,[38,39] etc. They show accuracy and transferability comparable to those of quantum-mechanical simulations but require less computational effort by many orders of magnitude. This enable computational simulations accessible to large systems at long timescale, applicable to phenomena such as phase transition[40] and crystallization[41].

Given the virtue in cost-accuracy tradeoff of ML potentials, they are promising substituent of DFT for accelerating the structure search. However, for a ML potential to be used for structure predictions, it is essential to take into account structures lying in different regions of the PES during the training process. This is different from most of previous works where only structures close to local minima of the PES are considered. Utilizing of ML potentials to accelerate structure predictions is now in its early stage but have already shown several encouraging results.[39,42,43]

In this work, we first demonstrate the applicability of current state-of-art ML potential, Gaussian approximation potential (GAP),[34] in reconstruction of the PES for global structure search. Then two efficient schemes have been developed for accelerating structure predictions by combining our in-house developed CALYPSO method with GAP. Boron (B) clusters were used as testing systems. Our newly developed schemes can readily reproduce previous experimental ground-state structures for $B_{36}$ and $B_{40}$ clusters with less computational cost by one order of magnitude compared with full DFT-based structure predictions. Furthermore, large-sized boron cluster containing 84 atoms are also studied, and putative global stable structure is predicted.

## 2. Method and Applications

### 2.1 The applicability of stat-of-art ML potential in reconstruction of PES for structure prediction

There are two central components of any ML potential: one is descriptor that represent a structure numerically in a unique and unambiguous way; the other is ML methods that reconstruct the PES using descriptor as input. Currently, plenty of descriptors are available for ML models, among which the atom centered symmetry functions (ACSF),[44] bispectrum of neighbor density and smooth overlap of atomic positions[45] have been demonstrated to be particularly suitable for fitting PES. For learning method, the high-dimensional neural network[46] and Gaussian process regression[34] show most promising solutions to this problem. In principle, one can choose from the above descriptors and ML methods arbitrarily to construct a ML potential. Here we employ the ACSF as descriptor, while the Gaussian process regression as implemented in the GAP is used for ML method.

ACSF descriptor[44] is a set of radial and angular functions which describe the coordination environment of an atom depending on the positions of the neighboring atoms up to a cutoff radius. An arbitrary number (typically 50-100) of symmetry function values for an atom *i* can be obtained by adjusting parameters in ACSF, which can be used to represent the environment of atom *i* in the sturcture and as input vectors $(d_i)$ for the ML model (Detail formulations of ACSF in shown in ESI).

Within the GAP frame,[34,35] the total energy of a system is described as a sum of atomic energies,

$$E = \sum_{i=1}^{N} \varepsilon_i(d_i),$$

where $\varepsilon_i$ is the *i*-th atomic energy. Atomic energies are interpolrated in the ACSF descriptor space through gaussian process regression, which is a kernal-based ML technique. The central results of GAP are predictions of atomic enegy ($\varepsilon_*$) and correspoinding variance ($\sigma_*$) described as the following two formulas,

$$\varepsilon_*(d_*) = k_*^T C^{-1} t,$$

$$\sigma_*(d_*) = \kappa - k_*^T C^{-1} k_*,$$

where $t$ is the vector of reference data, i.e., DFT total energes and atomic forces, $k_*$ is the covariance vector of functional values $k_* \equiv \langle \varepsilon_* t \rangle$, $C$ is the covariance matrix defined as $C \equiv \langle tt^T \rangle$, and $\kappa$ is the covariance between predicted atomic enegy and itself $\kappa \equiv \langle \varepsilon_* \varepsilon_* \rangle$. In practice, a sparse procedure is introduced to eliminate similar atomic configurations to improve the numerical stability and reduce the computaional cost.

In the current work, we further define the variance of total energy for a strucutre *u* as follows,

$$\Sigma_u = \frac{1}{N} \sum_{i \in u} \sigma_i(d_i),$$

where $N$ is the number of atoms in strucutre $u$. It is can be used as a metric for estimating the reliability of the predicted total energy.

Below we illustrate the performance of current GAP model in reconstruction the PES using B clusters as testing systems. The complex energy landscapes of B clusters are originated from the electron-deficient nature of B, leading to the occurrence a number of intriguing structures with very different geometries (e.g., planar,[47] tubular,[48] cage-like[49–51], bi-layered[52] and core-shell structures[53,54]).[55] This pose formidable challenges for ML potentials to reconstruct the PES.

We start with generating two reference datasets at DFT level as training and testing set, respectively (details for DFT calculations can be found in ESI). Training set contains 12053 data (structure and corresponding total energy as well as forces on each atom) corresponding to 357646 atomic environments. Testing set contains 3571 data with 102438 atomic environments. The data are obtained by local structural optimizations of a number of $B_n$ cluster randomly generated by CALYPSO in size range of n = 12-42. Since structures are randomly generated without any system-specific prior information, both datasets are expected to cover a wide range of the PES (geometrically and energetically) and contain various atomic environments.

Then we trained a series of GAPs by gradually enlarging the size of the training data from 100 to 12000. Two ACSF descriptors with 65 and 89 function values are used (denote as ACSF-65 and ACSF-89, the detailed parameters for the ACSF descriptors are listed in Table SI and SII in ESI). For each size of training data, ten potentials are generated with data randomly extracted from the training set. Fig. 1(a) and (b) show the evolution of the root mean square errors (RMSEs) for energy and force of these potentials tested on the testing set, respectively. It is clearly seen that the mean values of RMSEs of both energy and force significantly decrease as the size of training data increases, indicating gradual improvement of the potentials. The standard deviation of the RMESs is also decreased rapidly as the training size increases, indicating the performance of the potentials is gradually uncorrelated with specific choice of the training data. The potential obtained using ACSF-89 and 12000 training data gives the best performance, and comparisons between its predictions with DFT results for energy and force are showing in the insert of Fig. 1(a) and (b), respectively. The data points mainly distribute along the line with slope 45° corresponding to a perfect fit, leading to small RMSEs of energy and force (53 meV/atom and 430 meV/Å) which is comparable to that of original GAP for silicon clusters.[45] Note that the lengths of ACSF descriptors considered here (65 and 89) only have minor effects on the RMSEs of energy and force, but the longer ACSF descriptors (ACSF-89) gives better performance for potentials at large training size. We expect the RMSEs of energy and forces can be further decreased by increasing the training size and length of the ACSF descriptors but at the expense of higher computational cost during the training process.

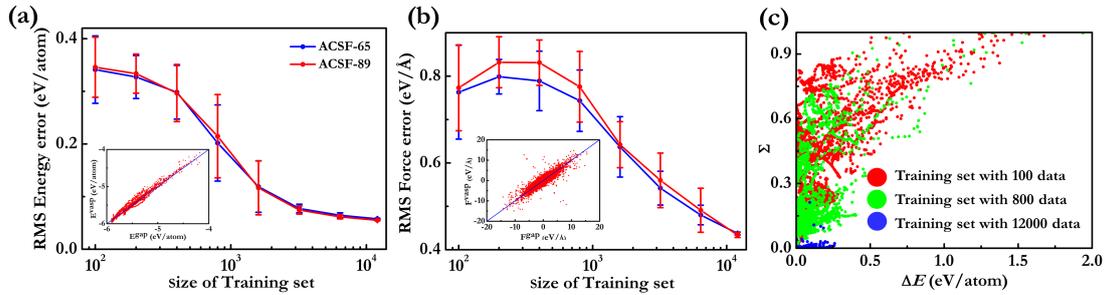

**Fig. 1** Evolution of RMSEs for GAP predicted energy (a) and force (b) on the testing set as a function of training size. Best GAP are obtained using ACSF-89 and 12000 training data, and comparisons between its predictions with DFT results for energy and force are shown in the inserts. (c) Correlations between energy error ($\Delta E$, energy difference between GAP and DFT) and predicted variance ($\Sigma$) for structures in the testing set. Data points for potentials using 100, 800 and 12000 training data are denoted as red, green and blue, respectively.

Fig. 1(c) shows the correlation between energy error ($\Delta E$, energy difference between GAP and DFT) and predicted variance ($\Sigma$) for structures in the testing set. As expected from the above results, both the extents of $\Delta E$ and $\Sigma$ decrease as training size increases from 100, 800, to 12000, indicating the improvement of the potential. Furthermore, the up-triangle $\Delta E - \Sigma$ distribution demonstrate that $\Sigma$ is indeed a viable indicator for the reliability of the predicted total energies of structures.

To further evaluate the transferability of the GAP, we performed local optimizations of 257 random structures using GAP and DFT, respectively. The distributions of interatomic distance and energy of optimized structure are shown in Figure 2(a) and (b). The distribution of interatomic distances for optimized structure using different methods are quite similar, with the first peak centered at ~ 1.65 Å. Energy distribution for GAP-optimized structures slight shift toward higher-energy regions compared with that of DFT, but the overall difference (88 meV/atom) is close to the RMSE of GAP (53 meV/atom). Visual inspection of optimized structures confirm that the two procedures generally give quit similar final configurations. Thus, the above results indeed suggest a high transferability of the current GAP model. This shed light on the potential of replacement of cost DFT calculations by GAP-based one for accelerating structure prediction.

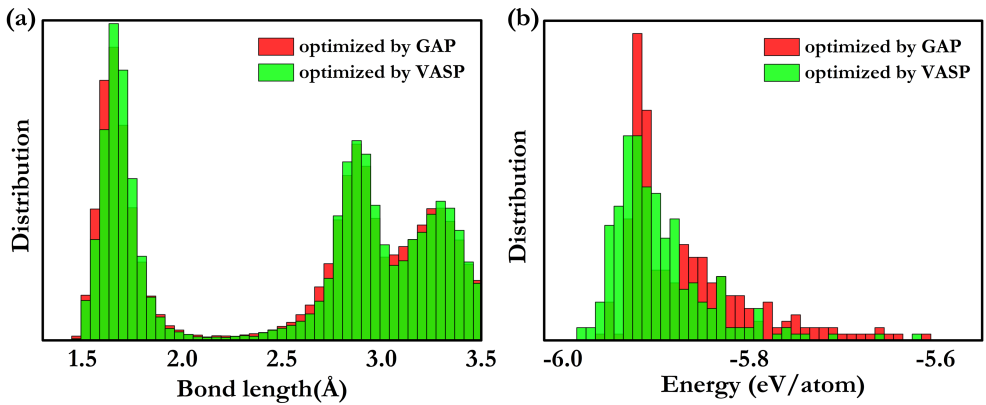

**Fig. 2** The distributions of interatomic distances (a) and energies (b) for 257 random structures of B cluster optimized using GAP and DFT.

## 2.2 Acceleration schemes for CALYPSO structure prediction

Encouraging by the above observation, we have developed two acceleration schemes for structure predictions. This is done on the top of our CALYPSO method,[13,14] which can explore the PES and locate stable/metastable structures intelligently by using the chemical composition alone for a material. Its validity and efficiency has been manifested by its successful applications in a variety of material systems at different dimensions,[15,16] such as 0D nanocluster,[50,56] 2D layered materials[57,58] and surface reconstruction,[59] as well as 3D crystal.[60,61] The key features of this methodology are swarm-intelligence based global optimization strategy and several particular devised structure dealing techniques. For more detailed theory of the CALYPSO method we refer the readers to Refs. [13,14,62–64] but give descriptions on its main procedure below.

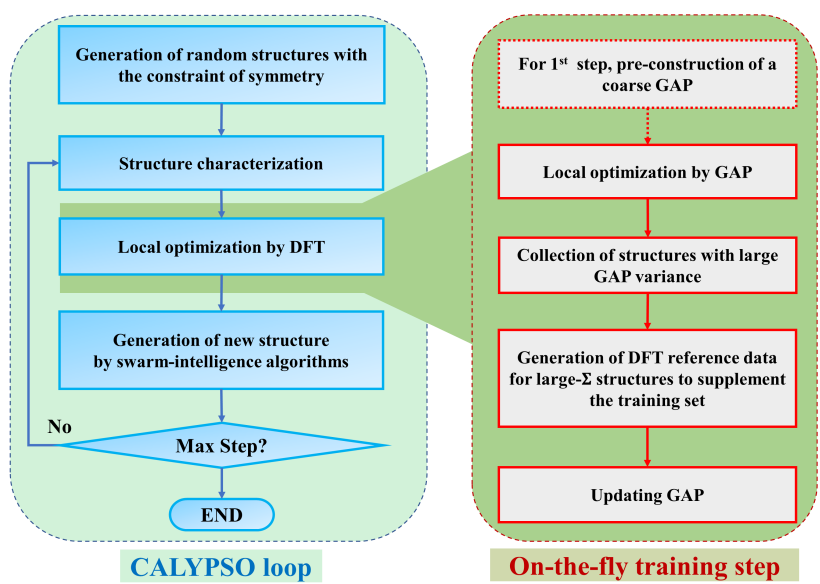

**Fig. 3** Flow chart of original CALYPSO structure searches (left panel) and acceleration scheme with on-the-fly training of GAP (right panel).

Structure prediction through CALYPSO comprises mainly four steps as depicted in the flow chart on the left panel of Fig. 3. First, the initial structures are randomly generated with the constraint of symmetry to allow a diverse sampling of PES. Then, bond characterization matrix[14] or coordination characterization function[65] techniques will be used to characterize the new structures and examine their distances (similarities) to all the previous ones. Structures with distances less than a threshold will be eliminated. After a user specified number of structures (a population or generation) have been generated, local structure optimizations based on DFT are performed to eliminate the noise of energy surface and drive the systems to the local minima. Eventually, swarm-intelligence algorithms (e.g., particle swarm optimization or symmetric artificial bee swarm) are applied to produce new structures for the next generation. This process continues iteratively until the maximum step is reached.

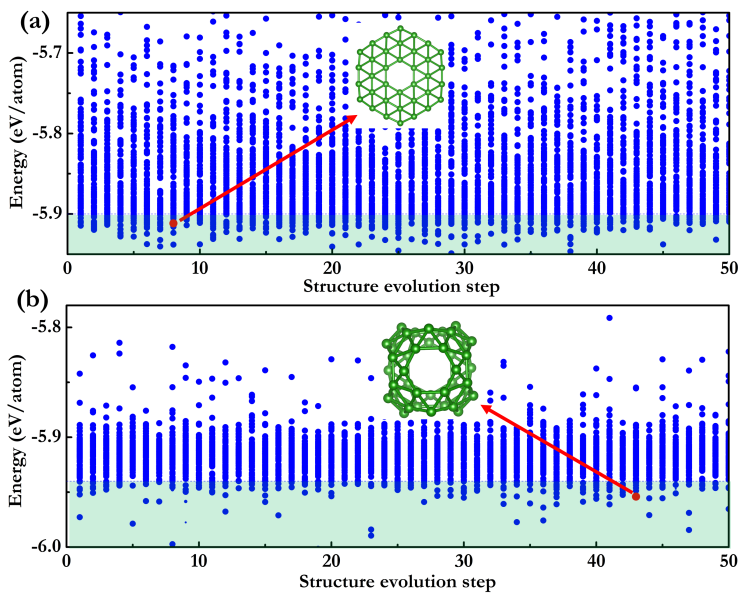

**Fig. 4** Evolution of energy distributions during CALYPSO structure search where local optimizations are fully conducted by GAP for $B_{36}$ (a) and $B_{40}$ (b) clusters.

In the first acceleration scheme, we just fully replace DFT calculations by GAP-based one in local structure optimizations. This scheme has been tested on predictions of structures of $B_{36}$ and $B_{40}$ clusters. The experimental planar $B_{36}$ and cage-like $B_{40}$ structures are intentionally removed from the training set during the construction of GAP. Evolutions of energy distribution during the CALYPSO structure searches are depicted in Fig. 4(a) and (b), respectively. It is remarkable that the two experimental structures can be readily reproduced during the searches. Although the two structures are not predicted as ground-state structures different from DFT results, they are lie in the low-energy regions below an energy threshold (defined as predicted lowest-energy for structures in the structure search plus the RMSE of the GAP), and this error can be easily remedy by subsequent DFT calculations of structures below the energy threshold (shaded areas in Fig. 4(a) and (b)). For each testing case, we stop the structure search at the 50$^{th}$ structure evolution step. About 300 DFT-based local optimizations were performed for training GAP and refining the low-lying structures below the energy threshold, while the other local optimizations are performed at GAP level, whose computational costs are negligible. For DFT-based structure predictions, 7000 local optimizations at DFT level are needed in each case. Therefore, such scheme can save the computational cost by at least 1 order of magnitude. The speed-up ratio can be even higher if longer structure evolutions are performed. However, in practice we usually carry out structure searches for a system from scratch, a well-trained GAP for the system under investigation is not exists. Below we develop an alternative scheme which can be accelerated by GAP and simultaneously train the GAP in an on-the-fly manner.

In the second acceleration scheme, we replace the step of DFT-based local optimizations by a process of on-the-fly training of GAP which is depicted on the right panel of Fig. 3. At the first step of CALYPSO structure search, an arbitrary GAP is provided. All subsequent local optimizations are performed at GAP level with final configurations subject to DFT single-point calculations. All structures during local optimizations are monitored, and the ones with predicted variance $\Sigma$ larger than a threshold are gradually accumulated. When a certain number of such structures are collected, DFT calculations are performed for these structures with data supplemented to the training set, and GAP is updated according to the new training set. This process ensures a gradual improvement of GAP by constructing the training set in an information-efficient way (i.e., avoiding the inclusion of similar structures to the training set). Moreover, as structure search proceed, progressively fewer training process and DFT calculations are needed.

We here illustrate the performance of this scheme through structure search of larger B cluster containing 84 atoms, whose structural elucidation is still elusive with both planar and core-shell structures are suggested to be the ground state.[66,67] In the first step of the CALYPSO search, a coarse GAP is trained from data obtained by DFT-based local optimization of a random structure of $B_{84}$ cluster. In the subsequent steps, all the structures are local optimized at GAP level, but the energy of final configurations are evaluated at DFT level. Structures with predicted variance $\Sigma$ larger than 0.03 are gradually collected, and DFT calculations are performed when

300 large-variance structures are obtained, whose data are supplemented to the training set to improve the GAP.

To evaluate the evolution of the GAP during the structure search, we construct a testing set containing 3500 random structures of $B_{84}$ clusters. Fig. 5(a) shows the evolution of RMSEs of GAP predicted energy and force for the testing set. At the first five GAP training steps, the RMSEs decrease rapidly. After that, the RMSEs gradually converge to ~65 meV/atom and 440 meV/Å for energy and force, respectively, which is very close to that of the best GAP obtain above. This maybe approach the best performance of current GAP model on description of PES for B clusters. Since B is a very challenging system, we expect much better performance of GAP for other simpler systems, such as metals. As the enlarging of the training set and improving of the GAP, more structure evolution steps are needed to collect sufficient number of large-variance structures (Fig. 5(a)). So progressively fewer DFT calculations are needed, and the frequency for training GAP decreases.

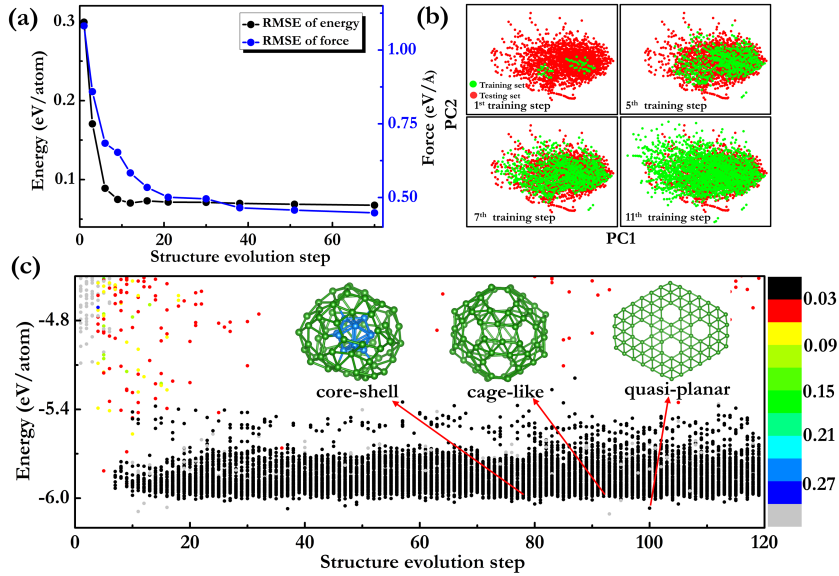

**Fig. 5** (a) Evolution of RMSEs of GAP predicted energy and force on the testing set during the CALYPSO structure search, (b) the PCA analysis of training set and testing set at the $1^{st}$, $5^{th}$, $7^{th}$, and $11^{th}$ GAP training steps, and (c) the evolution of the energy distributions during the CALYPSO structure search. Data points are shaded according to the value of predicted variances $\Sigma$. Lowest-energy structures for core-shell, quasi-planar and cage-like configurations are show in the inserts.

We performed principal component analysis (PCA) to examine the evolution of training set. In PCA, the ACSF descriptor for representing atomic environments is linearly transformed into a set of uncorrelated and orthogonal variables, known as principal components (PCs). The majority of the information contained within the ACSF descriptor can be captured by a few such PCs. We found that the first two PCs captures 98% information of the ACSF descriptor. Training data at the $1^{st}$, $5^{th}$, $7^{th}$ and $11^{th}$ training step are projected onto the 2-dimensioanl space of the first two PCs as shown in Fig. 5(b) compared with those of testing data. It is clear seen that the training data rapidly expand and gradually cover the testing set, indicating gradual completeness of the training set.

The evolution of the energy distributions during the structure search is shown in Fig. 5(c) with data points are colored according to its predict variance $\Sigma$. It can be seen that the predicted variances of structures are overall decrease during the structure search as a consequent of enlarging of the training set and improving of the GAP. The newly developed scheme allows to a fast sampling of about ~17000 local minimum structures, while only 2300 DFT single-point calculations are performed during the structure search. For *ab initio* structure prediction method, ~17000 DFT-based local optimizations (at least corresponding to $1.7 \times 10^6$ DFT single-point calculations) are needed. Thus, the acceleration scheme saves the computational cost by ~3 orders of magnitude compared with full DFT based structure search in this testing case.

The lowest-energy isomers for core-shell, quasi-planar and cage configurations in the current structure search are shown in the insert of Fig. 5 (c) (more low-lying isomers are given in Fig. S1 in ESI). The lowest-energy structure of quasi-planar configuration with four hexagonal holes is the same as that of recent theoretical study.[66] New structures for core-shell and cage configurations are predicted. The energies of these low-lying structures calculated at different level of theory at list in Table I. At PBE level, the quasi-planar structure possesses the lowest energy. But at PBE0 and TPSSh level of theory, our newly predicted core-shell structure is lower in energy than the quasi-planar structure by 27.5 and 15.2 meV/atom, respectively. Recent benchmark calculations indicate that TPSSh functional can give consistent results as that of high level theory of CCSD(T) for large-sized B clusters.[52] Thus, the core-shell structure are likely to the ground state for the $B_{84}$ cluster. In fact, our structure search calculations revealed a large number of low-lying structures with irregular core-shell configurations. This indicate a glasslike PES of $B_{84}$ dominated by core-shell configurations.[68] Thus, the core-shell structure is preferred for $B_{84}$ in both energetic and kinetic view.

**Table I.** Relative energies of the predicted core-shell, quasi-planar and cage-like structures at PBE/PW, PBE0/6-31G(d) and TPSSh/6-31G(d) level of theory. Energies are relative to the core-shell structure. Computational details can be found in ESI.

| meV/atom | core-shell | quasi-planar | cage-like |
| --- | --- | --- | --- |
| PBE/PW | 0.0 | -12.1 | -1.29 |
| PBE0/6-31G(d) | 0.0 | 27.5 | 29.0 |
| TPSSh/6-31G(d) | 0.0 | 15.2 | 28.1 |

## 3. Conclusions

In summary, using B clusters as testing systems, we have demonstrated the predictive capability of a well-trained GAP, which is qualified to be used to replace DFT during the global structure search. Then, two acceleration schemes for structure prediction of large systems have been developed by combining GAP with our CALYPSO structure prediction method. One can choice between the two schemes dependent on whether a well-trained GAP is available. Testing on medium- and large-sized B clusters demonstrated the efficiency and reliability of the current acceleration schemes. Moreover, putative global minimum structures of $B_{84}$ cluster are proposed. We adopted GAP and CALYPSO method in the current work, but the proposed schemes are general for current ML potentials and structure prediction methods. Thus, the current results represent a significant step toward structure prediction of large systems.

## ACKNOWLEDGMENTS

The authors acknowledge the funding support from National Natural Science Foundation of China (Grants Nos. 11534003, 11604117, 11404128 and 61775081), the Science Challenge Project, No. TZ2016001, National Key Research and Development Program of China (Grant No. 2016YFB0201200), and Program for JLU Science and Technology Innovative Research Team. Part of the calculations were performed in the high performance computing center of Jilin University and Tianhe2-JK in the Beijing Computational Science Research Center.